          [1999/12/02 1.01 (PWD)]
\documentclass[a4paper,twocolumn]{esapub} 
\usepackage{natbib}
\usepackage{epsfig}
\title{The Center of Our Galaxy:
Activity and High-Energy Emission of the Closest Massive Black Hole}
\author{A. Goldwurm}
\affil{Service d'Astrophysique /DAPNIA/DSM/CEA - Saclay, 
91191 Gif sur Yvette Cedex, France}

\begin{document}
\maketitle
\begin{abstract}
The Center of our Galaxy is a peculiar region where a number 
of crucial astrophysical phenomena take place, from star formation
to SN explosions and accretion onto a massive black hole.
The quest for a massive black hole in the Galactic Nucleus
is of course of particular relevance because, 
it would be the closest of such extreme objects, which are now 
believed to reside in most of the galactic nuclei of the universe.
I will review here the main observational characteristics of the 
Galactic Center with particular attention to the the problem of 
existence, physical condition and activity of the  
3 10$^6$~M$_{\odot}$ black hole coincident with the compact radio 
source Sgr A$^*$.
I will report historical and recent results of high energy 
observations of the central degree of our Galaxy, along with 
the specific accretion models proposed to account for the 
apparent lack of high energy activity from Sgr A$^*$.
The scientific perspectives of the next X and $\gamma$-ray missions
in the domain of the Galactic Center physics are also mentioned.
\end{abstract}
\section{Introduction: the Nuclear Bulge}
\begin{figure}
\centering
{\epsfig{figure=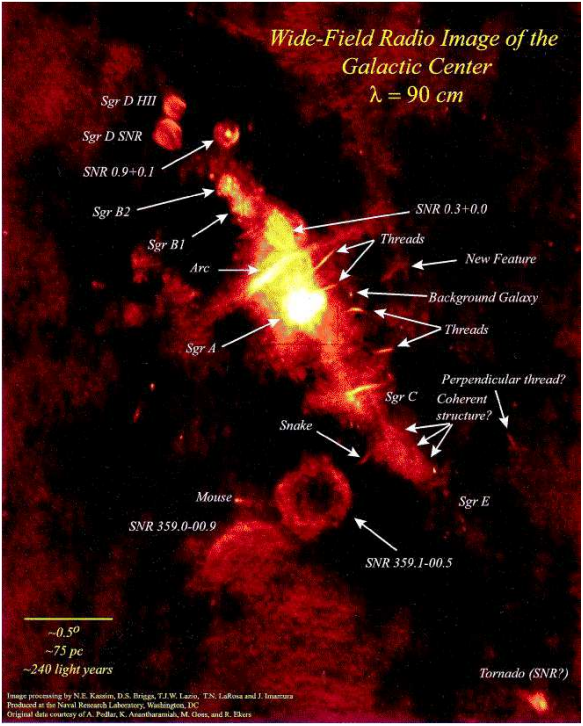, width=77mm}}
\caption{
Very Large Array (VLA) image at 90~cm of the Nuclear Bulge.
\label{fig:VLA90}}
\end{figure}

Already in the decade 1920-1930 
it was observed that the globular clusters are 
distributed with spherical symmetry around a point located in the 
Sagittarius constellation and also that stars were rotating
around the same point of the sky.
We now know that the dynamical center of the Galaxy is indeed located in
Sagittarius, at about 8 kpc from the Sun, 
right in the middle of the Milky Way. 
The Sun also lies on the galactic plane and the Galactic Center (GC) 
is highly absorbed by all the galactic disk dust which intercepts 
the line of sight.
Optical, UV and soft X-rays are therefore totally masked by an absorption
of A$_V$~$\approx$~31~mag corresponding to a column density
N$_H$~$\approx$~6~10$^{22}$~cm$^{-2}$, and the study of the
central regions can be carried out only from radio to near infrared 
(NIR) frequencies or at energies $>$~1~keV.
\\
The recently published VLA pictures at 90~cm \citep{LAR00} 
(Fig.~\ref{fig:VLA90}) 
show the richness and complexity of the Galactic 
Center region.
In the central 600 pc $\approx$~4$^{\circ}$ 
(at 8 kpc 1$''$~$\approx$~0.04 pc), 
a sky area often referred as the Nuclear Bulge,
the interstellar matter (ISM) is concentrated in a narrow layer (50~pc)
of molecular gas for a total mass of $\approx$~10$^8$~M$_{\odot}$.
Half of this gas forms dense, cold Giant Molecular Clouds (GMC), 
the denser of which are Sgr~B, Sgr~C, Sgr~D and those in the Sgr~A complex.
They have typical densities of 
$\approx$~10$^4$ -10$^5$~cm$^{-3}$ and temperatures of 30$-$200~K.
Several Supernova Remnants (SNR) are visible in the picture along with 
other filaments and structures of non-thermal emission, like the prominent 
Radio Arc, a filamentary source crossing the galactic plane.
\\ 
A number of thermal radio structures are also present, like the Bridge,
which is connecting the Arc to the Sgr~A complex. 
Some are identified with HII regions, like the features named
Pistol and Seiklo, whose gas is ionized by young hot stellar clusters,
like Quintuplet and Arches.
Magnetic fields are strong ($\approx$~2~mG) compared to typical values of 
the galactic disk ($\approx$~10~$\mu$G)
and lines are perpendicular to the galactic plane in the intercloud gas 
and parallel to the plane in the GMCs. 
Cloud kinematics indicates large velocity fields and a general inflow towards 
the center with rate of $\approx$~$10^{-2}$~M$_{\odot}$~yr$^{-1}$.
Stellar content of the Nuclear Bulge is basically an extension of the 
Galactic Bulge population.
It is dominated by a central nearly-isothermal 
cluster of low-intermediate mass stars, 
of metallicity $\approx$ 2, with density increasing towards the GC
with decreasing radius 
as ~R$^{-1.8}$, from R~$\approx$~100~pc down to a core radius 
R$_c$~$\approx$~0.1 pc where the star density reaches a constant 
value of $\approx$~10$^{7}$~M$_{\odot}$~pc$^{-3}$ and the mass 
a core value of M$_c \approx$~10$^{5}$~M$_{\odot}$. 
However, unlike the Galactic Bulge, this old-middle age (1-10~Gyr) 
population coexists with a young (10$^7$-10$^8$ yr) population 
of massive stars whose total fraction increases towards the center
and whose presence indicates recent star formation.
\\
In the following I will describe results obtained for the 
inner 50 pc and from high energy observations of the central degree.
For extensive reviews see Mezger et al. (1996) and 
Morris $\&$ Serabyn (1996), while for recent results see
Vol. 186 of ASP Conf. Series (eds. Falke et al. 1999) 
and Yusef-Zadeh et al. (2000).
%
\section{The ~Sgr~A~ Radio and Molecular Complex}
\begin{figure}
\centering
{\epsfig{figure=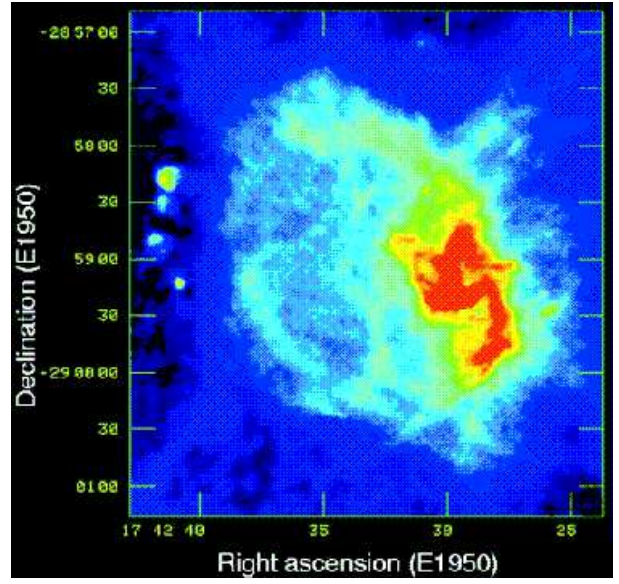 ,width=80mm}}
\caption{
VLA (6~cm) images of Sgr~A East halo (blue) and shell (light blue),
the inner thermal minispiral (red) and the 4 HII regions on the East side.
\label{fig:SgrAE}}
\end{figure}

\begin{figure}
\centering
{\epsfig{figure=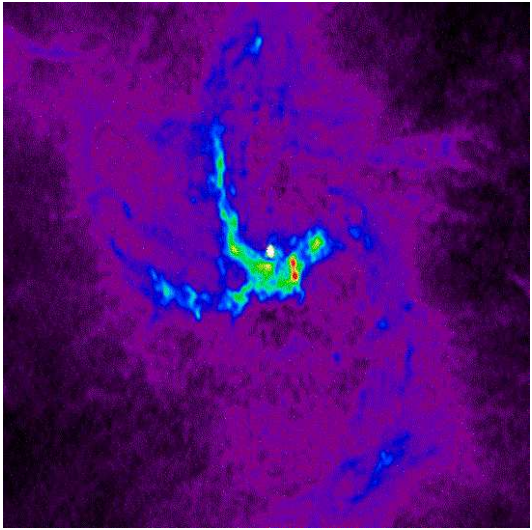,width=70mm}}
\caption{
Sgr~A West minispiral observed with the VLA at 1.2 cm.
The central white spot is Sgr~A$^*$.
\label{fig:SgrAW}}
\end{figure}

The central 50~pc (20$'$) are dominated by the Sgr~A radio complex
(Fig.~\ref{fig:SgrAE}, from Yusef-Zadeh et al. 2000). 
Relevant components of the Sgr A complex are, listing from the outside to 
the inner regions: 
several dense molecular clouds, the non-thermal Sgr~A East source, 
the CircumNuclear Disk (CND), the thermal source Sgr~A West 
and the compact source Sgr~A$^*$.
A recent study of molecular gas velocities of the region \citep{COI00}
describes the morphology and distribution of matter of the complex. 
Two dense molecular clouds account for most of the ISM of the region, 
interact with, and probably supply matter to the inner central region.
M-0.02-0.07, also known as the 50 km s$^{-1}$ cloud, 
lies on the north-east side
of the GC and includes the Sgr~A East Core very dense MC 
(15~pc size and 2~10$^{5}$~M$_{\odot}$ mass) 
observed to surround Sgr~A East from behind the GC. It is connected by
a molecular ridge to the North side of the other cloud, M-0.13-0.08 
(the 20 km s$^{-1}$ cloud), which is located South-East of the GC, 
about 10 pc in front of it and which
seems to supply the CND of molecular gas through the so called ``southern 
streamer''.
M-0.02-0.07 and its molecular ridge are compressed by the expanding shell 
of Sgr~A East. Sgr~A East is a non-thermal source composed by a diffuse
halo of triangular shape (7$'$ $\times$ 10$'$) 
and an oval shell (7 pc $\times$ 9 pc i.e.
3$'~\times$~4$'$) with major axis parallel to the galactic plane and
centered about 50$''$ ($\approx$ 2~pc) west of Sgr~A$^*$.
The shell appears in expansion, compressing the cloud and probably 
creating the string of 4 HII regions at the border with M-0.02-0.07 
and the OH masers also observed around the shell.
A life time of 5 10$^{4}$ - 10$^{5}$ yr and a total explosion energy of
4~10$^{52}$~ergs were derived from the source radio brightness and 
expansion velocity.
40 nearly simultaneous SN are needed to account for this energy 
and other scenarios
have been proposed to explain Sgr A East energetics, including the
explosive tidal disruption of a 1~M$_{\odot}$ star by the central massive 
black hole. 
Between 1.7$-$7 pc (40$''$-175$''$) from the center lies the CND
a clumsy, asymmetrical torus of neutral molecular gas and dust
rotating (V$_R \approx$~100~km~s$^{-1}$) around Sgr~A West.
The CND mass is 10$^4$~M$_{\odot}$, the gas temperature $\geq$100 K
and it is observed in molecular transition lines (e.g. HCN) 
and also in infrared 
wavelengths, emitted by the dust heated by radiation coming
from the central cavity.
Its inner part has the shape of a ring whose edges bound 
the central cavity and the thermal source Sgr~A West. 
The HII region Sgr~A West was unambiguously separated by the non-thermal
emission of Sgr~A East in 1975, while the inner minispiral pattern
was discovered in 1983 with the VLA. 
The bulk of the ionized gas (250~M$_{\odot}$) appears as
extended emission of size 2.1~pc $\times$ 2.9~pc with average density 
n$_e$~$\approx$~10$^3$~cm$^{-3}$, in which is embedded the denser
{\it minispiral}, a 3 armed spiral structure (Fig.~\ref{fig:SgrAW}) 
orbiting around the GC, composed by northern and eastern arms, 
a central bar and a western arc.
The radio spectrum resembles that of a typical optically thin HII region 
with T$_e$~$\approx$~6000~K, 
emission measure $\approx$~2~10$^3$~pc~cm$^{-6}$ 
and ionization temperature T~$\approx$~3~10$^4$~K.
These parameters imply the presence of a ionizing source of UV luminosity 
L$_{UV}$~$\approx$~7.5~10$^{37}$~erg~s$^{-1}$ 
and total flux of Lyman continum photons
N$_{Lyc}$~$\approx$~1.2~10$^{50}$~ph~s$^{-1}$.
Recombination emission lines allowed detailed kinematic studies of the gas,
which show that Sgr~A West is rotating from north-east to south-west and then
towards north-west. 
This seems to imply that the western arc is the ionized edge 
of the CND, the N and E arms are tidally stretched streams of infalling gas 
and the bar is an extension of the N arm.
Sgr~A West is seen in front of Sgr~A East and Sgr~A East Core cloud 
but is behind the 20~km s$^{-1}$ molecular cloud.  
Little north of the minispiral central bar and visible as a white point in  
Fig.~\ref{fig:SgrAW} lies the compact source Sgr~A$^*$ which must be seen
in front of the minispiral bar even if embedded in the diffuse ionized cloud.
\section{The Compact Radio Source Sgr~A$^*$}
\begin{figure}
\centering
{\epsfig{figure=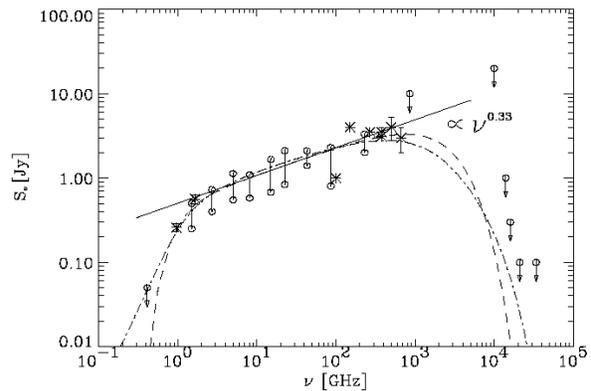 ,width=80mm}}
\caption{
Radio spectrum of Sgr~A$^*$. Minimum and maximum fluxes 
show the variability level.
\label{fig:SgrA*}}
\end{figure}

Sgr~A$^*$ was discovered by Balick $\&$ Brown (1974), 
3 years after Lynden-Bell $\&$ Rees (1971) 
had predicted that a compact synchrotron radio source should
reveal the presence of a massive black hole (MBH) in the Galactic Nucleus.
Sgr~A$^*$ is indeed a compact, bright, non-thermal radio source, 
which coincides (within 50~mas) \cite{GHE00} 
with the dynamical center of the Galaxy.
The radio spectrum (Fig.~\ref{fig:SgrA*}) is an inverted power-law
($S_{\nu} \div \nu^{\alpha}$) 
with spectral index $\alpha~\approx~0.33$ between 1~GHz
and 800~GHz, and with low and high frequency cut-offs. 
Flux variability of 30-100$\%$,
around a value of $\approx$~1~Jy,
on timescales of few months is observed,
and the average radio luminosity is estimated to $\approx$~300~L$_{\odot}$.
The source is also very static, an upper limit of 20~km~s$^{-1}$ 
has been set to its proper motion. 
Considering the high velocities of the stars of the region
(500-1000 km s$^{-1}$) this unusual low value indicates that Sgr~A$^*$
must be massive. A lower limit of 
1000~M$_{\odot}$ to its mass was indeed derived, 
excluding the possibility of a stellar object.
The apparent radio size of Sgr~A$^*$ increases with wavelength
as $\lambda^2$,
which is the sign of source broadening due to ISM electron scattering.
However at low enough $\lambda$ ($<$ 7 mm) the relation seems to flatten 
and the source is probably resolved by the VLBI. 
The most recent VLBI data at 1.4~mm provide a size of 
$\approx$~0.1~mas \citep{KRI98}
which at 8 kpc corresponds to only 1.2~10$^{13}$~cm = 0.8~AU or,
in terms of a 3~M$_{\odot}$ BH Schwarzschild radius (see $\S$~6), 
of only $\approx$~14~R$_S$.
Moreover the source appears elongated with the intrinsic
major axis $\approx$~3 times greater 
than the minor axis and oriented in the North-South direction \citep{LO98}.
This result, if confirmed, could indicate the presence of a weak radio-jet.
Circular polarization is observed at several radio frequencies but 
until recently only upper limits ($<$ few $\%$) were set for linear 
polarization. 
Aitken et al. (2000) have now discovered linear polarization at
level of 10-20~$\%$ at sub-mm frequencies ($\geq$~100~GHz).
The sub-mm part of the Sgr~A$^*$ spectrum seems peculiar 
also because it shows a bump of emission with respect to the 
extrapolation of the radio power-law. Indeed,
using simultaneous observations at different frequencies
from 10~cm to 1~mm, Falke et al. (1998) found at $>$~100GHz an 
excess which implies a change in the spectral slope. 
\section{Evidences for a Massive Black Hole at the GC}
\begin{figure}
\centering
{\epsfig{figure=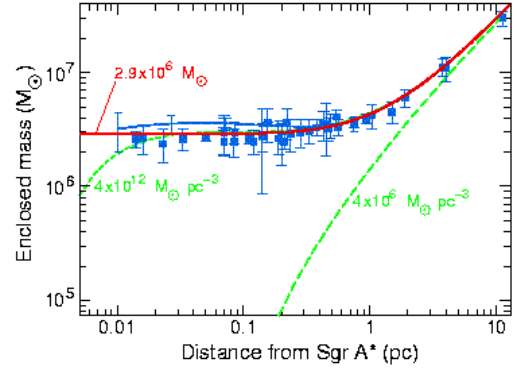 ,width=80mm}}
\caption{
Enclosed mass vs. distance from GC.
\label{fig:EncMass}}
\end{figure}

High gas velocities in the vicinity of the putative MBH of the
galaxy were also predicted by Lynden-Bell $\&$ Rees in 1971 
as a consequence of the gravitational pull exerted by the BH. 
These authors suggested to search for radio recombination lines 
of the thermal gas close to the BH to measure gas velocities
and then to estimate the BH mass.
High velocities of the Sgr~A West ionized gas have been indeed observed,
since 1974, using the H recombination lines (H109$\alpha$, 
H91$\alpha$, H92$\alpha$) the forbidden line $\lambda$12.8~$\mu$m
of [NeII] and the $\lambda$2.17~$\mu$m Br$\gamma$ line 
\citep{MEZ96, YUS00}. These measures could show that the
dynamics at radii R~$>$~1.5-2 pc (i.e. outside the CND inner ring)
is dominated by the nearly-isothermal central star cluster, while within
1.5~pc from the center the dynamics is due to a central mass of 
$\approx$~3~10$^6$~M$_{\odot}$ enclosed within $\approx$~0.17 pc. 
However gas motion measures cannot constrain further the volume of 
the central mass and also non-gravitational forces 
(magnetic fields, turbulence, etc.) may play a role
making the interpretation of the recorded velocities not obvious.
\\
Star kinematic studies, based on both stellar dispersion and 
rotation velocities, started in 1978, with the aim of evaluating
the central mass M(R) enclosed within the radius R down to distances 
$<$~0.1 pc. 
The most convincing results have been obtained 
in the last 8 years thank to high resolution NIR observations
performed with speckle/adaptive optics in the K band (2.2$\mu$m),
which allowed to provide independent determination of
star velocities using measures of star proper motions.
The recent paper by Genzel (2000) summarizes the results obtained both 
with the SHARP CCD camera mounted on the 3.5m ESO NTT telescope
(Genzel et al. 1997) and with the NIRC camera at the 10~m Keck telescope 
(Ghez et al. 1998).
About 1000 star with m(K) $<$ 16 (= 0.25 mJy) were imaged between
1992-1999 within the central 1~pc (25$''$) 
with angular resolutions in the range 0.15-0.05$''$.
Proper motions for 100 stars 
could be determined along with 200 high quality spectra.
The important improvement provided by proper motion measures 
is that they allow 
measurements of stellar velocities of the very faint stars which are 
very close (R~$<$~0.6$''$) to Sgr~A$^*$ and which show motions at 
$>$ 1000 km s$^{-1}$ (1470 km s$^{-1}$ for the closest star 
in projection, at 0.1$''$ = 800~AU from the radio source).
The simultaneous measure of proper motion and radial velocity for 32 stars 
between 1$''$-5$''$ from Sgr~A$^*$ also allowed to test the fundamental 
hypothesis of velocity isotropy assumed by most of the mass estimators.
Combining radial and proper motions data and using various projected mass
estimators Genzel et al. (2000) derived the mass distribution reported
in Fig.~\ref{fig:EncMass}.
The enclosed mass for R~$>$~1-2~pc can be accounted for by the isothermal
star cluster (broken line in Fig. 5) 
but at lower radii, in particular to fit data 
at 0.015~pc ($\approx$ 3000 AU), 
it is necessary either to assume a point mass of 
2.9~10$^6$~M$_{\odot}$ (full line) 
or to invoke the presence of a dark cluster of 
objects with central density of 4~10$^{12}$~M$_{\odot}$ pc$^{-3}$
(dotted line).
Even composed of stellar BHs, such a cluster would not be stable 
for more than 10$^{7}$ yr and the hypothesis of a massive black hole
at the Galactic Nucleus is now extremely strong.
The very recent discovery with the Keck of curvature in the trajectories 
of 3 of the closest stars to Sgr~A$^*$ \citep{GHE00}, 
confirms that stars are orbiting around the GC, increases by a factor 10
the density required for an alternative dark cluster
and proves that the radio source is at less than 0.05$''$ 
from the dynamical center of the stars.
\\
NIR observations have also led to the discovery of a cluster of about 
25 bright, young, hot stars most of them belonging to 
the complex IRS 16 centered $\approx$~2$''$ east of the GC. 
These very luminous stars show broadened and 
P-Cyg type HeI/HI emission lines which indicate that their helium rich 
surfaces are expanding as powerful stellar winds with velocities of
V$_{_W}$$\approx$ 500-1000 km s$^{-1}$ and mass loss rates 
$\dot M_{_W} \approx$ 1-80 10$^{-5}$~M$_{\odot}$~yr$^{-1}$ \citep{NAJ97}. 
These outflows produce a hypersonic wind of density 
n$_{_W}$$\approx$ 10$^{3}-$10$^{4}$ cm$^{-3}$ 
which pervades the central pc. 
These stars also emit $\approx$ 10$^{38}$~erg~s$^{-1}$ 
in UV and can therefore totally account for the entire UV luminosity 
required to excite Sgr~A West and to heat the dust of the central cavity 
and of the CND ($\S$~2).
To conclude with IR observations we have to mention the important upper limits
set by Menten et al. (1997) on the NIR flux from Sgr~A$^*$ which imply
an L$_{NIR}$~$<$~10$^{35}$ erg~s$^{-1}$ for the compact radio source.
\section{High Energy Observations of Sgr~A$^*$}
\begin{figure}
\centering
{\epsfig{figure=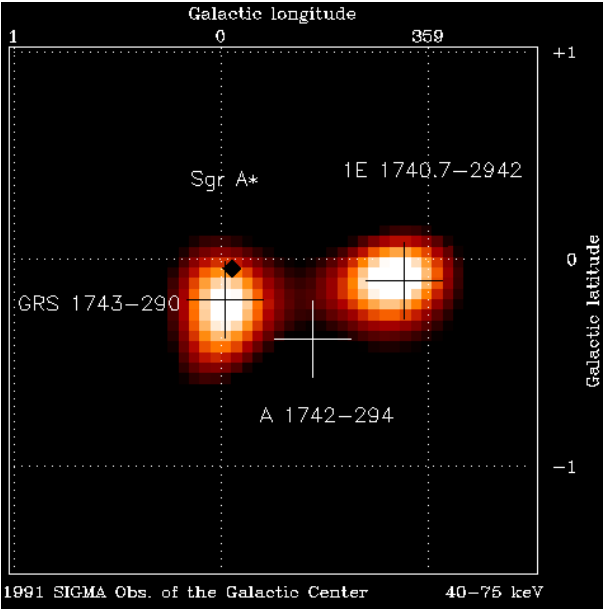 ,width=80mm}}
\caption{
Image around Galactic Nucleus obtained by SIGMA in 1991. 
Black diamond is position of Sgr~A$^*$, and crosses of 
SIGMA sources of the region.\label{fig:SigmaGN}}
\end{figure}

After the first detections of high energy emission from the general direction
of the GC made with non-imaging instruments in the decade 1970-1980,
the first real X-ray images with arcmin resolution were obtained 
with the Einstein Observatory \citep{WAT81}. 
In addition to diffuse emission, 12 point sources were detected within 
the central 20$'$, one of which, 1E~1742.5$-$2859, observed at $<$~1$'$ from
Sgr~A$^*$ with L$_{1-4~keV}$~=~1.5~10$^{35}$~erg~s$^{-1}$.
Between 1980-1990 several observations were carried out
in the standard and hard X-ray bands and a number of transient sources
were detected.
The most remarkable observations were certainly those performed
with the coded mask XRT instrument on SpaceLab2, which provided the first
ever images (3$'$ resolution) of the GC in hard X-rays (3-30~keV) 
\citep{SKI87}. 
Sgr~A$^*$ appeared again rather faint 
(L$_{3-30~keV}$~=~5~10$^{35}$~erg~s$^{-1}$),
in spite of the high expectations for an hard spectrum of the 
type of Cyg~X-1.
However in 1991 the Rosat telescope with 25$''$ resolution in the 
band 0.1-2.5~keV, could separate 
1E~1742.5$-$285 in 3 sources one of which, RXJ~1745.6$-$2900,
was found within 10$''$ from Sgr~A$^*$ \citep{PRE94}. 
Rosat luminosity (L$_{0.8-2.5~keV}$~=~3~10$^{34}$~erg~s$^{-1}$)
however falls below 2 orders of magnitude
from extrapolation of the N$_H$-corrected XRT luminosity. 
The possibility of strong variability or additional N$_H$ were 
considered but ASCA 1993-1994 observations in the range 0.1-10 keV, 
revealed presence of 2 point sources, 
a soft and stable one close to the Nucleus and another harder and transient
about 1$'$ away \citep{KOY96}. The authors concluded then that
the flux observed by the hard X-ray instrument on SL2 
(also confirmed by ART-P/GRANAT, Pavlinsky et al. 1994) 
was due to the transient source and not to Sgr~A$^*$.
\\
ASCA also confirmed and improved the results of EXOSAT,
Tenma and Ginga on the presence of hard X-ray diffuse emission. 
A large part of the Nuclear Bulge seems permeated by hot plasma
emitting X-ray diffuse emission with thermal spectrum of T~$\approx$~10~keV 
and numerous K emission lines of He-like and H-like ions. 
The diffuse source has an elliptical shape 1$^{\circ}$~$\times$~1.8$^{\circ}$ 
elongated on the galactic plane. The emission peaks around the center
where a hot spot with an oval 2$'~\times$~3$'$ shape 
(= 4.8 pc $\times$ 7.2pc)  
and luminosity of $\approx$~10$^{36}$~erg~s$^{-1}$ was also imaged with ASCA.
This emission is puzzling because at this temperatures the gas would not 
be bound by the GC gravitational potential. 
The estimated expansion of the large shell
at sound speed provides an age of $\approx$~50000~yr and input energy of 
$\approx$~10$^{54}$~erg for the gas, and continuous heating is required
with power of $\approx$~10$^{41-42}$~erg~s$^{-1}$. 
Tanaka et al. (2000) have recently argued against thermal origin of this
emission and proposed that charge-exchange interaction of 
low-energy cosmic-rays with ISM contribute to it.
ASCA also detected 6.4~keV neutral iron line diffuse emission in 
the Nuclear Bulge, and more precisely from Sgr~B2 the most
massive of the GMC. This fluoroscence line can be produced by 
high energy emission which is scattered in the neutral environment of 
a GMC. 
Murakami et al. (2000) estimated, comparing data to simulations,
that an external source at a distance d, should have emitted about
L$_{2-10~keV}$~=~3~10$^{39}$~(d/100 pc)$^2$~erg~s$^{-1}$ over 100 yr.
This excludes binary transient sources but leaves open the 
possibility for a flare of Sgr~A$^*$.
If emitted from Sgr~A$^*$, radiation should have been travelling through 
$\approx$ 300 l.y. to reach SGR B2. These data could then probe 
300 yr ago flaring activity from Sgr~A$^*$. 
Similar conclusions were reached by Sunyaev $\&$ Churazov (1998)
who investigated also dependence of the line characteristics 
and variability with position and time behavior of primary source,
while Fromerth et al. (2001) found that data are also
compatible with location of primary source inside the GCM.
\\
To a be a faint source in X-rays does not prevent to be
strong emitters in gamma rays, in particular for
such extreme objects like BH, which are known to emit very hard radiation
and suspected to generate e$^+$-e$^-$ annihilation line. 
The GC has been a priority target of the first satellite imager 
of soft $\gamma$-rays (30 keV - 1300 keV) SIGMA. SIGMA with its 13$'$-15$'$ 
resolution and large field of view performed deep surveys of the galactic 
bulge between 1990-1997 cumulating nearly 10$^{7}$~s of data.
SIGMA found that in this band the 1$^{\circ}$ region around 
the GC is dominated by the otherwise anodyne
X-ray source 1E 1740.7-2942, 
which follow up observations revealed to be associated to radio-jets 
and which is now one of the 4 persistent galactic BH candidates X-ray sources.
Sgr~A$^*$ appeared instead silent. 
The only weak flare observed by SIGMA from the vicinity of the GC in 1991
(see Fig.~\ref{fig:SigmaGN}),
was attributed to a point source $>$~9$'$ away from Sgr~A* \citep{GOLD94, VAR96}.
A recent re-analysis of the whole SIGMA data by Goldoni et al. (1999)
have provided 
the best low-energy $\gamma$-ray upper limits for this source
(see also Fig.~\ref{fig:IBISsen} and Goldwurm et al. 2000). 
The results imply L$_{30-300~keV}$~$<$~1.2~10$^{36}$~erg~s$^{-1}$,
and an upper limit (3~$\sigma$) of 3.3~10$^{-4}$~ph~cm$^{-2}$~s$^{-1}$ 
for narrow 511 keV line point source at the GC 
was also obtained with SIGMA \citep{MAL95}.
OSSE/GRO mapped the galactic diffuse 511 keV emission 
and could identify a bulge component of 
about the same value in flux than the above quoted SIGMA upper 
limit, with a positronium fraction of 0.89 \citep{PUR97}.
Due to the limited angular resolution ($>$~4$^{\circ}$), 
however it was not possible to estimate if any of this emission 
is from a central point source.
\\
A source (2EG~J1746-2852) in the Galactic Center
was also detected by EGRET/GRO \citep{MAY98} at $>$ 30 MeV. 
The poor angular resolution of the instrument ($\sim$1$^{\circ}$) 
does not exclude it is actually diffuse. 
The spectrum is a broken power-law with photon indexes of 1.3 and 3.1 
below and above 1900 MeV, which seems against
pure $\pi_0$ decay origin, with total luminosity of 
L$_{>100 MeV}$ = 2.2~10$^{37}$~erg~s$^{-1}$. 
SAX (NFC/MEC) monitoring of GC 
has not provided new results on the Nucleus itself, 
however it was found that the soft component (T~$\approx$~0.6 keV) 
of the local diffuse emission seems related to Sgr A East and it is well 
interpreted as thermal emission from a SNR \citep{SID99}. 
The hard component (T~$\approx$~8 keV) instead  
lies along the plane as shown previously by ART-P and ASCA.
\\
Chandra observations with the unprecedent angular resolution 
of ~0.5$''$, carried out in fall 1999, have instead 
resolved the Rosat source in few components, one of which 
could be point-like and lies within 0.35$''$ from Sgr~A$^*$ \citep{BAG01}.
The source, which may well be the real X-ray counterpart of the
radio source, has a power-law spectrum rather steep ($\approx 2.7$) 
and luminosity L$_{2-10~keV}$~$\approx$~2~10$^{33}$~erg~s$^{-1}$.
Althought the 0.5-2 keV flux is very uncertain due to the low 
accuracy of N$_H$ estimation it seems clear that Sgr~A$^*$ soft X-ray 
luminosity is well below the Rosat value.
\section{The Problem of Sgr~A$^*$ Low-Luminosity}
In spite of the very compelling evidences for a MBH at the Galactic Nucleus,
Sgr~A$^*$ does not behave like a simple mass scaled-down AGN.
Summing measured luminosities and considering the quoted upper 
limits, Sgr~A$^*$ bolometric luminosity is 
$$ L_{Sgr~A^*}~<~5~10^{36}~erg~s^{-1}$$
The UV luminosity is uncertain due to the high extinction but the fact that
Sgr~A West ionization can be provided by the IRS~16 cluster 
of hot stars implies
that there is not an additional strong source of UV, whose emission
would also be re-emitted by the heated dust and observed in IR.
The EGRET source can rise this value by a factor 
5 but in any case the Nucleus remains sub-luminous.
\\
For a BH at the GC with mass M = 3.0~10$^{6}$~M$_{\odot}$, 
the  Schwarzshild radius R$_S$ and Eddington luminosity L$_E$ are
$$ R_S = {2 G M \over c^2} = 9.0~10^{11} cm = 0.060~AU \sim 7.5~\mu as$$
$$ L_E = {4~G M m_p \over \sigma_T } = 3.9~10~^{44}~erg~s^{-1} $$
In spherical accretion, matter with density n$_{_W}$ 
flowing with relative velocity V$_{_W}$ 
within a distance R$_A$ (accretion radius) from the BH is captured, 
undergoes a shock which dissipates the kinetic energy in thermal energy   
and then is accreted radially 
with free-fall velocity and mass accretion rate 
$\dot M_A$ providing a power L$_A$ (accretion luminosity) given by
$$ R_{A} = {2 G M \over V_{_W}^2} = 1.6~10^{17}~cm = 0.05~pc \sim 1.3''$$
$$ \dot M_{A} =  R_{A}^2 \rho_{_W} V_{_W} = 5.5~10^{22}~g~s^{-1} =
8.7~10^{-4}~M_{\odot}~yr^{-1}$$
$$ L_{A} = {G M \dot M_{A} \over R_S} = 
2.4~10^{43}~erg~s^{-1} = 0.061~L_E$$
where wind parameters quoted in $\S$~4 where used,
i.e. V$_{_W}$ = 700 km s$^{-1}$, n$_{_W}$ = 5.5~10$^3$ cm$^{-3}$.
3D simulations of the stellar winds from the IRS 16 cluster 
flowing into the GC black hole have
been performed by Cocker and Melia (1997) using results of Najarro et al.
(1997) and they obtained a mass accretion $\approx$ 1-2 times
the value of $\dot M_A$. 
So the accretion luminosity is 10$^{6}$-10$^{7}$ 
times higher than the measured bolometric luminosity 
and efficiency of conversion of available energy 
in radiation must be lower than 10$^{-6}$.
The final efficiency depends actually on how the matter, once captured 
at R$_A$ is accreted, how reaches the BH event horizon, which are 
the characteristics of the flow and whether a disk forms.
Cocker and Melia (1997) in their simulations 
also found that the accreted specific angular momentum l 
in units of cR$_S$ ($\lambda$ = l/cR$_S$) varies around an average
of $\lambda$ = 40~$\pm$~10 with the sign swapping on 
time-scales of $\sim$~100 years.
This implies that the circularization radius R$_c~\approx~2\lambda^2R_S$
(distance at which angular momentum equals the Keplerian value) 
is $<$~3000~R$_S$ and a large accretion disk probably does not form.
\section{Accretion Models for the GC Black Hole}
\begin{figure}
\centering
{\epsfig{figure=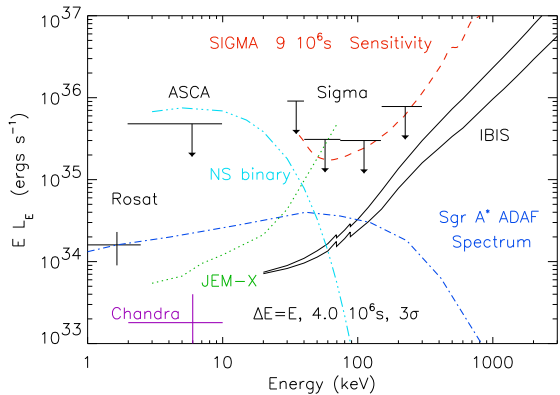, width=80mm}}
\caption{
IBIS (full lines) and JEM-X (dotted) INTEGRAL sensitivities and 
the ADAF spectrum (broken-dotted) of Sgr A$^*$ with
SIGMA and ASCA upper limits, and Rosat and Chandra luminosities.
\label{fig:IBISsen}}
\end{figure}

An optically thick and geometrically thin accretion disk 
has efficiency of the order of 0.1 and, if present around Sgr~A$^*$,
would originate a bright spectral peak in the UV band 
similar to the blue bump of AGNs,
followed by a steep Wien tail which may contribute little to X-rays. 
The low energy tail would extend to NIR and such component would 
violate the Menten et al. (1997) upper limits.
\\
The first attempt to fit the entire spectrum of Sgr~A$^*$ modeling
the accretion into the massive BH was done by 
Melia (1992, 1994), who computed emission produced 
in pure spherical Bondi-Hoyle geometry.
Without the viscous dissipation which occurs in disks, matter would 
fall into the hole heated up by simple adiabatic compression,
and efficiency of conversion of gravitational energy in radiation 
would be very small.
However it is usually assumed that matter carries, in the fall, the 
magnetic field, maintained in equipartition with the gas,
and magnetic dissipation by turbulence or field lines reconnection
can heat efficiently the particles which then
emit synchrotron, free-free and inverse Compton radiation. 
With this model Melia (1992) could fit the available data for a 
black hole mass of 10$^6$~M$_{\odot}$, 
a value smaller than recent estimates. 
In its 1994 model, Melia also included an optically thick
disk at small radii, to account for the effects of the accreted angular 
momentum, but the Menten et al. (1997) IR upper limits are 
not compatible even with such small disk.
Recently, Cocker and Melia (2000) revisited the model 
improving computation of the emitted spectrum and
assuming magnetic field B in sub-equipartition by letting 
its value to be free parameter. Considering variability the 
model luminosities seem marginally compatible with the data
and imply that B, which is of few mG at R$_A$ ($\sim$ 10$^5$~R$_S$),
does not reach the equipartition value at intermediate radii.
Only in the inner region of $\approx$~5-25~R$_S$, where 
gas circularizes and a small Keplerian disk is formed, 
the magnetic dynamo can rise the field to values of the order of 200 G.
The emission from this inner region could then account for the 
sub-mm excess and for its polarization. However this variant of the 
model fits the data only for an effective accretion rate 
$<$~10$^{-4}$~$\dot M_A$. 
\\
In 1995, Narayan et al. proposed that accretion flows
for sub-Eddington rates ($\dot M < \dot M_E$) 
in systems with BHs be dominated by advection. 
These models (ADAF) assume a very weak coupling between protons and electrons
in the flow and the energy carried by protons is not transmitted 
efficiently to the radiating electrons but rather advected into the hole.
\\
The plasma gets very hot (optically thin), flows with 2 temperatures
(kT$_p~\approx$~10$^9$~K, kT$_e~\approx$~10$^7$~K), 
with nearly spherical geometry (geometrically thick disks), 
but with viscous dissipation and momentum transport.
The radio spectrum is generated by optically thick self-absorbed
synchrotron emission 
from hot nearly-relativistic thermal electrons at different temperatures.
Each disk ring at radius R gives rise to a spectrum peaked at the 
critical frequency above which radiation becomes optically thin.
A total flat spectrum is then produced by the superposition 
of these ring spectra.
Highest radio frequencies are generated by the hottest electrons of the 
inner regions close to the BH while inverse Compton produces a weak
tail in IR, optical and UV, and X-rays are produced by thin thermal 
bremsstrahlung from electrons at all radii till R$_A$.
The best ADAF model for Sgr~A$^*$ \citep{NAR98} fits the  
radio data, the Rosat X-ray flux and the IR and hard X-rays upper limits 
(see Fig.~\ref{fig:IBISsen}) 
for the correct BH mass but for an accretion rate  
$\dot M \approx$ 10$^{-5}$~M$_{\odot}$ yr$^{-1}$. 
This is more than a factor 10
lower than estimated from the IRS~16 stellar winds.
Since in ADAF L~$\div$~$\dot M^2$ the luminosity for 10 times the rate
would be 100 times greater and the discrepancy is not easily explained 
\citep{QUA99b}.
To account for the EGRET $\gamma$-ray source Narayan et al. (1998) computed
the contribution of decay of 
pions produced by proton-proton collisions in the very hot inner region of 
the ADAF. The spectrum could reproduce the EGRET spectral shape but not the 
flux normalization.
On the other hands electrons produced by the muons seem able to naturally
account for the {\it cm} part of the spectrum 
not well explained by the thermal electrons \citep{MAH98}.
\\
The spectrum of Narayan et al. 1998 fitted the Rosat flux 
and the new value of X-ray luminosity obtained with Chandra certainly
makes more acute the problem of accretion rate (see Fig.~\ref{fig:IBISsen}).
Blandford $\&$ Begelman (1999) argued that ADAFs must have 
outflows in form of winds and, more recently, it was also found 
that convection in the disk may reduce outward transport of 
angular momentum reducing the net flow of matter into the hole 
\citep{QUA00}.
Both these new variants of ADAF (known as ADIOS and CDAF) could 
reduce effective mass accretion.
However Quataert $\&$ Narayan (1999) discussed the spectral shape 
of ADAF with winds and found that since soft X-rays
are produce far from the hole (between 10$^4$-10$^5$~R$_S$),
outflows, expected in the inner regions, will not reduce the
expected X-ray emission and discrepancy with the accretion rates
is not resolved.
Moreover the Chandra steep spectrum does not seem
compatible with thermal bremsstrahlung.
Another important difficulty for ADAF models is the recent discovery of linear 
polarization in the sub-mm band. In both ADAF and Bondi-Hoyle models
this excess is produced in the inner regions of the flow and therefore
even if radiation can be initially polarized, the accreting plasma 
encountered on the way out would depolarize the emission 
by Faraday effect \citep{AGO00}, unless very peculiar geometries are invoked.
Observed values of polarization seem to imply accretion rates as low
as 10$^{-8}$~M$_{\odot}$~yr$^{-1}$ making of course need for low-efficiency
models much less important.
\\
All these factors certainly revive the non-thermal models of Sgr~A$^*$.
The flat (index $\approx$ 0.3) radio spectrum could 
indeed be interpreted as due to thin synchrotron radiation 
from a quasi mono-energetic distribution of electrons, 
e.g. accelerated by shocks inside the accretion flow.
The low frequency cut-off is due to self absorption while the high frequency
one to the cut off in the truncated power-law electron distribution.
Such model was investigate by Beckert et al. (1996) and refined by
Beckert $\&$ Duschl (1997).
X-rays can then be produced by self synchrotron Compton (SSC)
emission. 
The main difficulty was to correctly account for the X-ray flux
and to explain the ``ad-hoc'' cut off in the electron distribution.
Another non-thermal model for Sgr~A$^*$ is the jet/nozzle model,
recently re-discussed by Falcke $\&$ Markoff (2000) (see references therein).
The authors propose that radio emission is generated 
in a compact radio jet, similar to those seen in AGNs, but less powerful. 
This model could explain the possible
asymmetry in the radio and sub-mm shape of the source \citep{LO98} 
which, if confirmed, would indeed favor the presence of a jet.
The base of the jet where the acceleration 
of the magnetized plasma takes place, the nozzle, 
is a compact region where the sub-mm radiation is originated. 
The nozzle is close to the BH and to the inner part of the accretion flow.
The jet mechanism is not known but the authors invoke a
coupling between jet and accretion disk.
As for the mono-energetic electron model the X-rays are produced
by SSC and the expected spectrum seems to fit well the recent Chandra
results.
However the accretion flow must not provide relevant X-ray  
emission and the problems of low-efficiency accretion or 
of a much lower effective accretion rate than the one estimated from
data on stellar winds remain.
\\
Finally, the Sgr~B2 6.4 keV line could indicate that Sgr~A$^*$ is flaring, 
may be cyclically like the X-ray Novae. 
However if the accreted material is stored between the flares, 
in a large low-viscosity disk, the latter should be visible in IR.
\section{Perspectives and Conclusions}
ADAFs have become popular in recent yeas because they seemed 
able to explain, in the frame of BH accretion disk thermal models, 
the low efficiencies observed in the nucleus of our Galaxy 
and of other close normal galaxies. 
However these models are now encountering serious difficulties, 
and non-thermal models become again competitive.
All models however fail to explain why the closest known 
massive black hole to us radiates so little, unless we admit 
that it does not accrete the available matter as expected.
Major advances in understanding of the physics of the Galactic Center
will be provided in the next years by new and more accurate observations,
in particular at high energies.
The Chandra and XMM-Newton observatories will probably address
the question of the X-ray (0.1-15 keV) luminosity, variability 
and spectrum of Sgr~A$^*$ and 
of the characteristics and origin of the local X-ray diffuse emission. 
The next ESA $\gamma$-ray (3 keV - 10 MeV) 
mission INTEGRAL will allow to study, if present, 
the ADAF hard X-ray component predicted by Narayan et al. (1998). 
Fig.~\ref{fig:IBISsen} reports the expected sensitivity 
of the medium energy (20 keV - 10 MeV) instrument (IBIS) onboard INTEGRAL
compared with the ADAF model. 
Detailed simulations \citep{GOLD00} show that
IBIS will be able to detect the expected thermal bremsstrahlung 
hard component of Sgr~A$^*$ in 4~10$^6$~s. 
With INTEGRAL it will be possible to search for a
compact source of 511 keV line at the GC. In spite most models 
do not predict a strong e$^-$-e$^+$ annihilation line contribution 
from the MBH itself, Fatuzzo et al. (2000) argued 
that part of the 511 keV bulge component may be due to cumulative
effect of pair production in the expanding shells of repeated
Sgr~A~East type events.
High energy observations with GLAST at $>$ 100 MeV will
resolve the standing problem of the EGRET source, 
possibly linked to the Sgr~A$^*$ or to the non-thermal Sgr~A East shell.
The latter is also expected to provide detectable flux in the TeV range,
making the GC an interesting target also for next generation of ground
VHE/UHE $\gamma$-ray and neutrino experiments.
Moreover the Galactic Center has been recognized as possible
site of large cold dark matter concentration and 
annihilation \citep{G&S99}.
\\
Sgr~A$^*$ is probably the closest massive black hole, and
in spite presently in low state of activity, 
remains a highly privileged target for astronomical multi-$\lambda$ 
observations and for future astroparticle experiments.
\section*{Acknowledgments}
I thank Fulvio Melia for very useful discussions and Frederick
K. Baganoff for providing information about Chandra 
observations before publication.
\end{document}